\documentclass[preprint]{revtex4}

\usepackage{color,psfig,graphics,psfrag,amsmath,amssymb,amsfonts,rotating}

\newcommand{\eps}{\varepsilon}
\newcommand{\s}{\sigma}

\begin{document}

\title{Message passing algorithms for non-linear nodes and data compression}

\author{Stefano Ciliberti}
\email{ciliberti@roma1.infn.it}
\affiliation{Laboratoire de Physique Th\'eorique et Mod\`eles Statistiques,
Universit\'e de Paris-Sud, b\^atiment 100, 91405, Orsay Cedex, France}

\author{Marc M\'ezard}
\email{mezard@lptms.u-psud.fr}
\affiliation{Laboratoire de Physique Th\'eorique et Mod\`eles Statistiques,
Universit\'e de Paris-Sud, b\^atiment 100, 91405, Orsay Cedex, France}

\author{Riccardo Zecchina}
\email{zecchina@ictp.trieste.it}
\affiliation{ICTP, Strada Costiera 11, I-34100 Trieste, Italy}

\begin{abstract}
The use of parity-check gates in information theory has proved to be very
efficient. In particular, error correcting codes based on parity checks over
low-density graphs show excellent performances. Another basic issue of
information theory, namely data compression, can be addressed in a similar
way by a kind of dual approach. The theoretical performance of such a Parity
Source Coder can attain the optimal limit predicted by the general
rate-distortion theory. However, in order to turn this approach into an
efficient compression code (with fast encoding/decoding algorithms) one must
depart from parity checks and use some general random gates. By taking
advantage of analytical approaches from the statistical physics of
disordered systems and SP-like message passing algorithms, we construct a
compressor based on low-density non-linear gates with a very good
theoretical {\it and} practical performance.
\end{abstract}

\maketitle

\section{Introduction}

Information theory and statistical physics have common roots. In the recent
years, this interconnection has found further confirmation in the analysis
of modern error correcting codes, known as Low Density Parity Check (LDPC)
codes~\cite{Gallager,Spielman,mackay}, by statistical physics
methods~\cite{Sourlas,montanari,nishimori}. In such codes, the choice of the
encoding schemes maps into a graphical model:  the way LDPC codes exploit
redundancy is by adding bits which are bound to satisfy some random sparse
linear set of equations -- the so called parity checks. Such equations are
used in the decoding phase to reconstruct the original codeword (and hence
the message) from the corrupted signal.  The parity checks can be
represented by a graph indicating which variables participate in each
check. The randomness and the sparsity of the parity checks is reflected
into the characteristics of the associated graphs, a fact that makes
mean-field type statistical physics methods directly applicable.  Thanks to
the duality between channel coding and source coding~\cite{mackay}, similar
constructions can be used to perform both lossless and lossy data
compression~\cite{jelinek,yedidia}. It must be mentioned that while there
exist some practical algorithms for the lossless source coding which are
very efficient~\cite{lossless}, much less work has been done so far for the
lossy case~\cite{bergergibson}. In this work we present a new coding
technique which consists in a generalization of parity-check codes for lossy
data compression to non-linear codes~\cite{cimeze}. This issue has been
addressed recently using methods from the statistical physics of disordered
systems: a non-linear perceptron~\cite{perceptron} or a satisfiability
problem~\cite{battaglia} have been used to develop practical source coding
schemes. Other recent advances in the field can be found
in~\cite{garciazhao,caire,yamamoto}.

We consider the following problem. We have a random string of uncorrelated
bits $x_1,\ldots x_M$ (prob$(x_a=0)=$ prob$(x_a=1)=1/2$) and we want to
compress it to the shorter coded sequence $\s_1,\ldots\s_N$ (encoding). The
rate $R$ of the process is $R=N/M$. Once we recover the message (decoding),
we may have done some errors and thus we are left in principle with a
different sequence $x_1^*,\ldots x_M^*$. The number of different bits
normalized by the length of the string is the distortion,
\begin{equation}
  D=\frac 1M \sum_{a=1}^M \left(  1-\delta(x_a,x^*_a)\right) \ .
\end{equation}
The Shannon theorem states that the minimum rate at which we can achieve a
given distortion $D$ is $R_{min}=1-H_2(D)$, where $H_2(D)$ is the binary
entropy of the probability distribution used to generate the $x_a$'s. This
$R_{min}$ is the value one has to compare to in order to measure the
performance of a compression algorithm. This problem is a simple version of
the lossy data compression problem. In real applications one is often more
interested in compression through quantization of a signal with letters in a
larger alphabet (or continuous signal~\cite{quantization}). Here we shall
keep to the memoryless case of uncorrelated unbiased binary sources, and we
hope to be able to generalize the present approach to more realistic cases
in the near future.

\subsection{Constraint Satisfaction Problems} 

\begin{figure}
  \psfrag{a1}[][][2]{$a_1$}
  \psfrag{a0}[][][2]{$a_2$}
  \psfrag{a3}[][][2]{$a_3$}
  \psfrag{a4}[][][2]{$a_4$}
  \psfrag{a5}[][][2]{$a_5$}
  \psfrag{v1}[][][2]{$\s_1$}
  \psfrag{v2}[][][2]{$\s_2$}
  \psfrag{v3}[][][2]{$\s_3$}
  \psfrag{v4}[][][2]{$\s_4$}
  \psfrag{v5}[][][2]{$\s_5$}
  \psfrag{v6}[][][2]{$\s_6$}
  \psfrag{v7}[][][2]{$\s_7$}
  \psfrag{b1}[][][2]{$b_1$}
  \psfrag{b2}[][][2]{$b_2$}
  \psfrag{b3}[][][2]{$b_3$}
  \psfrag{b4}[][][2]{$b_4$}
  \psfrag{a}[][][2]{$a$}
  \psfrag{0}[][][2]{$\s_0$}
  \psfrag{1}[][][2]{$\s_1$}
  \psfrag{2}[][][2]{$\s_2$}
  \psfrag{3}[][][2]{$\s_3$}
  \psfrag{4}[][][2]{$\s_4$}
  \psfrag{5}[][][2]{$\s_5$}
  \centering
  \includegraphics[angle=0,width=.45\columnwidth]{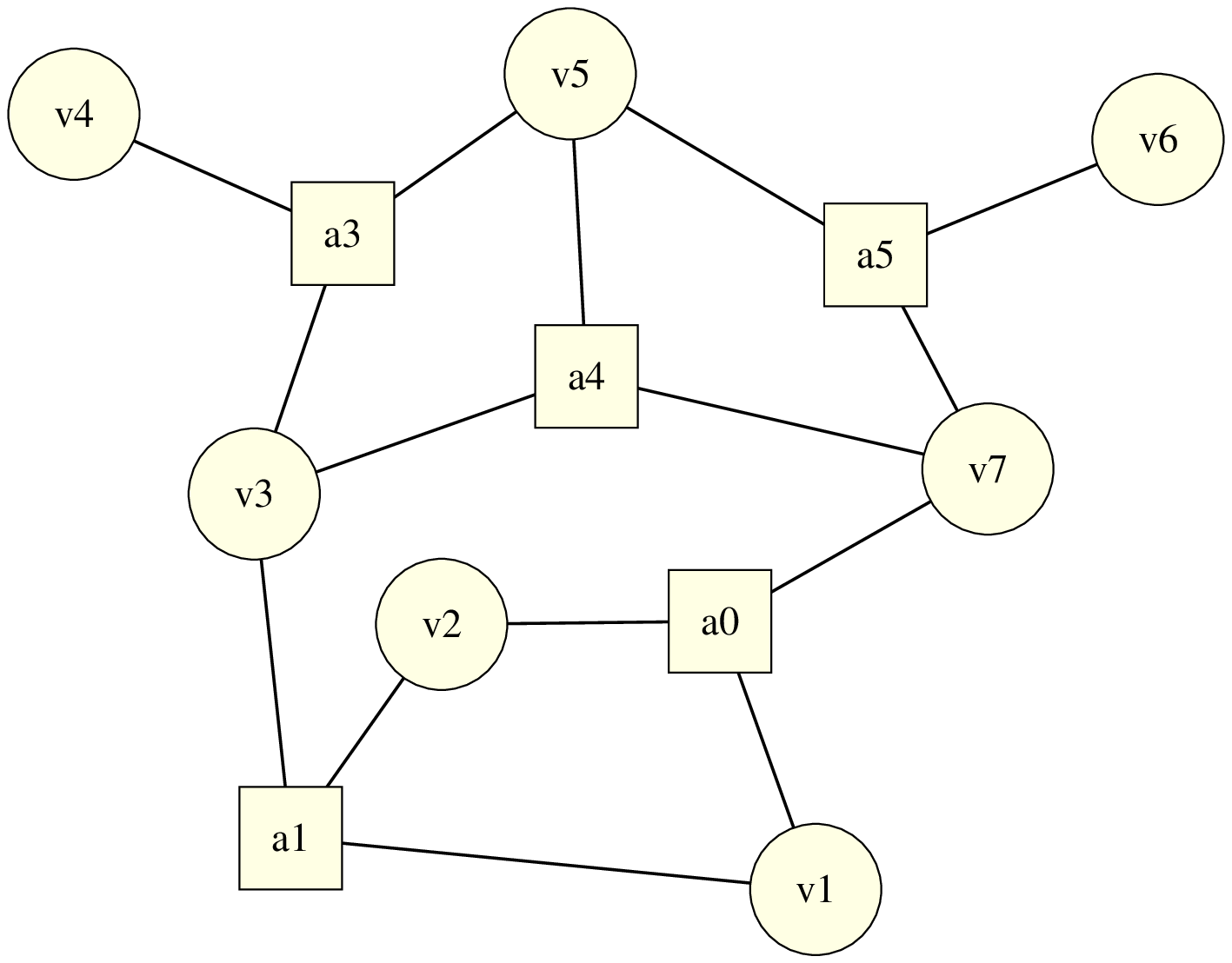}
  \includegraphics[angle=0,width=.45\columnwidth]{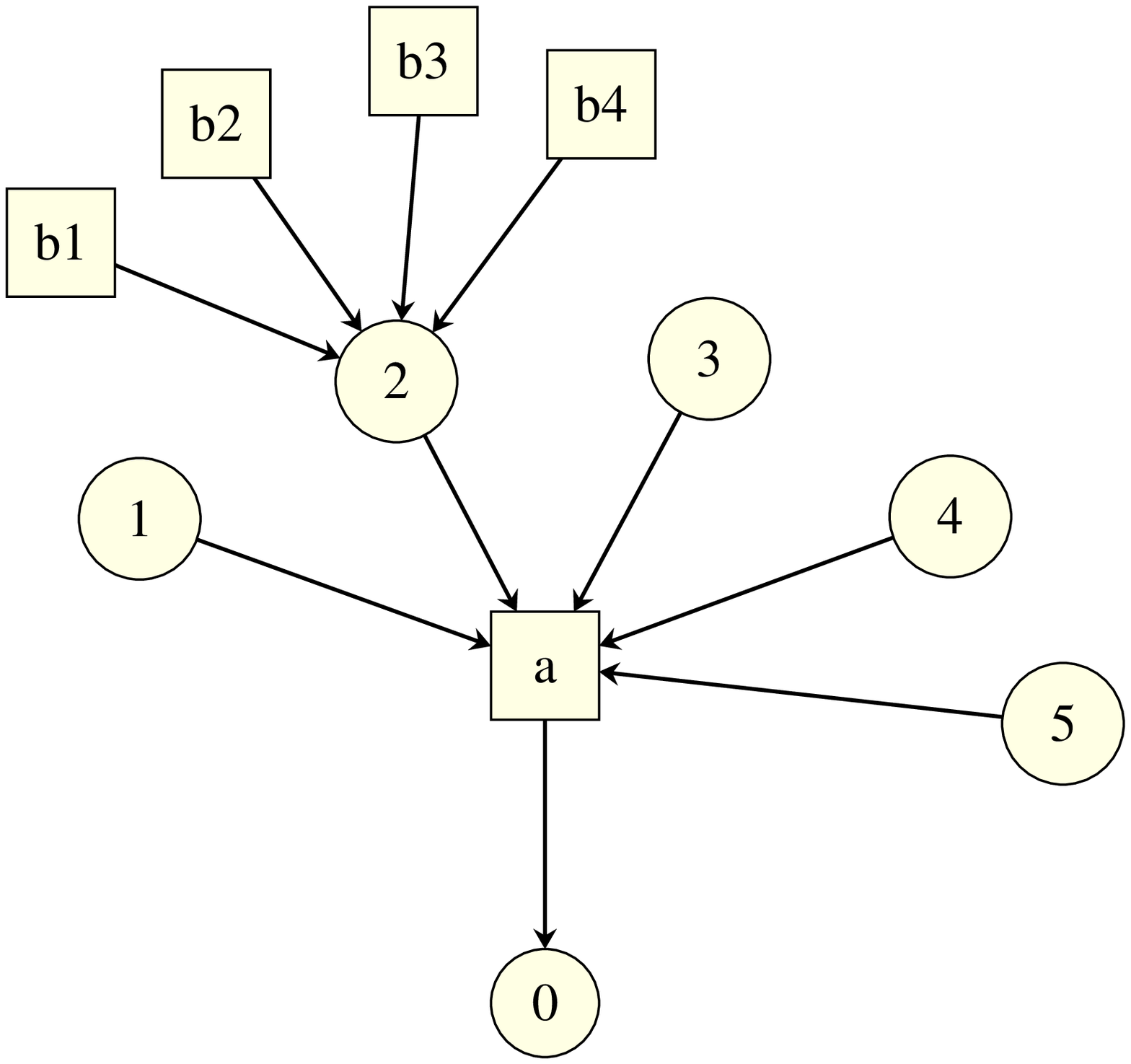}
  \caption{Left: Factor graph for a constraint satisfaction problem with
  $N=7$ variables (circles) and $M=5$ constraints (squares). In this example
  the connectivity of the function nodes is kept fixed, that is $K_a=K$ in
  the notations of the text. Right: Message passing procedure ($K=6$): The
  probability $q_{a\to 0}(u_{a\to 0})$ depends on all the probabilities
  $q_{b_i\to i}(u_{b_i\to i})$, $i=1,\ldots 5$ (see text).}
  \label{fig:factor}
\end{figure}

The underlining structure of the protocol that we are going to introduce is
provided by a constraint satisfaction problem (CSP). A CSP with $N$
variables is defined in general by a cost function
\begin{equation}
  E[\s] = \sum_{a=1}^M \eps_a\left(\{\s_i\in a\}\right) \ ,
\end{equation}
where the {\it function node} $\eps_a(\s_1,\ldots \s_{K_a})$ involves a
subset of $K_a$ variables randomly chosen among the whole set. The problem
is fully defined by choosing some $\eps_a$ and an instance is specified
by the actual variables involved in each constraint. A useful representation
for a CSP is the factor graph (Fig.~\ref{fig:factor}, left). The two kinds
of nodes in this graph are the variables and the constraints. For the sake
of simplicity we shall take the functions $\eps_a$ to have values in
$\{0,2\}$ (the clause is resp. SAT or UNSAT) and the variables $\s_i$ will
be Ising spins ($\s_i=\pm 1$). A somewhat special case of CSP which has been
studied extensively occurs when $K_a=K$ for any $a$, and we shall restrict
ourselves to this case in what follows. One can then show that the
degree of a given variable is Poisson distributed with mean value $KM/N$
and that the typical length of a loop is $\mathcal{O}(\log N
/\log(KM/N))$. We are interested in the limit $M,N \to \infty$, where
$\alpha\equiv M/N$ is fixed and plays the role of a control parameter.
Once a problem is given in terms of its graph representation, an
instance corresponds to a given graph chosen among all the legal ones with
uniform probability. We will go back to this problem in section
\ref{sec:formalism}. We first show how a CSP can be used as a tool for
compressing data.

\subsection{Encoding}

We use the initial word $x_1,x_2,\ldots x_M$ to construct a set of $M$
constraints between $N$ boolean variables $\s_1,\s_2,\ldots \s_N$. The
factor graph is supposed to be given. Each constraint $a$ involves $K_a$
variables and is defined by two complementary subsets of configurations
$\mathcal{S}^a_0$ and $\mathcal{S}^a_1$. The value of $x_a$ controls the
role of the subsets as follows: If $x_a=0$ then all the configurations
$\{\s_1\ldots\s_{K^a}\}\in\mathcal{S}^a_0$ satisfy the constraint, {\it
i.e.} $\eps_a=0$; all the configurations
$\{\s_1\ldots\s_{K^a}\}\in\mathcal{S}^a_1$ do not satisfy the constraint and
thus $\eps_a=2$. If $x_a=1$ the role of the two subsets is exchanged.  This
defines the CSP completely and we can look for a configuration of $\s_i$
that minimizes the number of violated constraints. This is the encoded word
and the rate of the process is $R=1/\alpha$.

\subsection{Decoding}

Given the configuration $\s_1,\s_2,\ldots \s_N$, we compute for each node
$a$ whether the variables $\s_{i_1^a},\ldots \s_{i_{K_a}^a}$ are in
$\mathcal{S}_0$ (in this case we set $x^*_a=0$) or in $\mathcal{S}_1$
(leading to $x^*_a=1$). Since a cost $\eps=2$ is paid for each UNSAT clause,
the energy is related to the distortion by
\begin{equation}
  \label{eq:distortion}
  D=\frac E{2M} = \frac E{2\alpha N} = R \frac E{2N}\ .
\end{equation}

\subsection{The Parity Source Coder}

\begin{figure}
  \centering
  \includegraphics[angle=-90,width=.8\columnwidth]{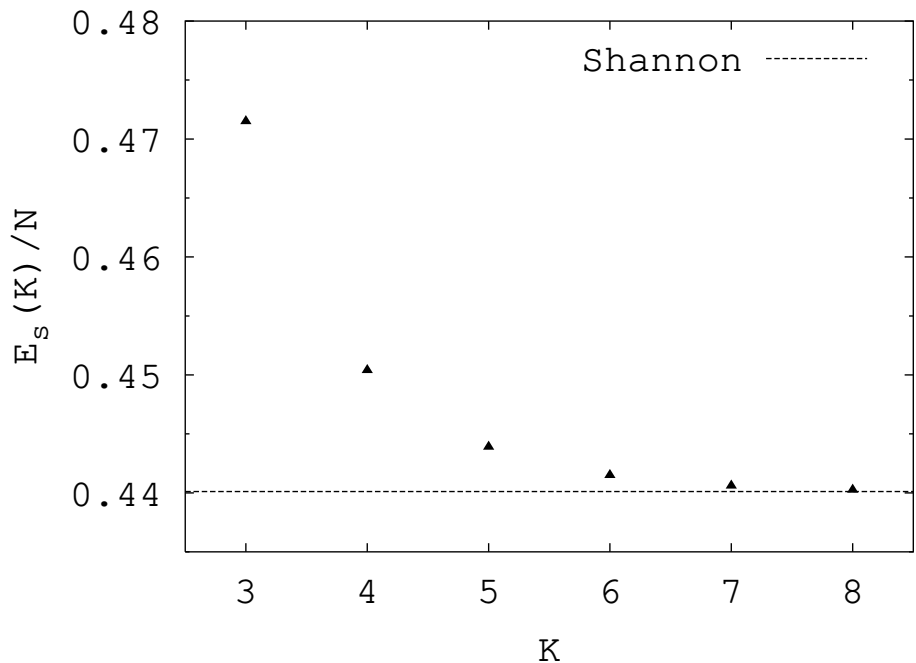}
  \caption{The ground state energy density for the XORSAT problem at $\alpha=2.0$
    versus $K$. We also show the corresponding Shannon bound as computed
    from rate-distortion theory, $E_{Sh} = 2\alpha D_{Sh}$, where $D_{Sh}$ is
    such that $1-H_2(D_{Sh})= 1/\alpha$.}
  \label{fig:exor}
\end{figure}

In order to give an example, we apply these ideas to the case where the CSP
is a parity check (the optimization problem is then called
XORSAT). In other words, the cost of the constraint $a$ is written as
\begin{equation}
  \eps_a =  1 +(-1)^{x_a} \prod_{i\in a} \s_i   \ .
\end{equation}
This Parity Source Coder (PSC) is the counterpart of the LDPC codes in
error-correcting codes and then it seems natural to expect a good
performance from it.

In Fig.~\ref{fig:exor} we show that the ground state energy of the XORSAT
problem, that is the theoretical capacity of the PSC, quickly approaches the
Shannon bound as $K$ increases~\cite{cime}. A PSC with checks of degree
$K\gtrsim 6$ has then a theoretical capacity close to the optimal one. The
problem here is that there does not exist any fast algorithm that can encode
a string (for example, the SID algorithm that we discuss in the next section
is known not to converge).

We will thus investigate a new kind of constraint satisfaction problem,
using non-linear nodes, with (almost) the same good theoretical performance
as the XORSAT problem, but with a fast encoding algorithm. Before we do
this, we introduce in the next section the general formalism used in the
study of non-linear nodes.

\section{The general formalism} \label{sec:formalism}

Given a CSP at some $\alpha$, we are interested in knowing whether a
\emph{typical} instance of the problem is satisfiable (i.e. all the
constraints can be satisfied by one global configuration) as the number of
variables goes to infinity. Generally speaking, there will be a phase
transition between a SAT regime (at low $\alpha$ the problem has a small
number of constraints and thus is solvable -- at least in principle ) and an
UNSAT regime where there are too many constraints and one cannot find a
zero-energy configuration. In this UNSAT regime, one wants to minimize the number
of violated constraints. This kind of problem can be approached by message
passing algorithms.

Due to the large length of typical loops, the local structure of a typical
graph is equivalent to a tree, and this is crucial in what follows. We
introduce the {\em cavity bias} $u_{a\to i} \in \{-1,0,+1\}$ sent from a
node $a$ to a variable $i$. A non-zero message means that the variable $i$
is requested to assume the actual value of $u_{a\to i}$ in order to satisfy
the clause $a$. If $u_{a\to i}=0$ the variable $i$ is free to assume any
value. It is clear that this message sent to $i$ should encode the
information that $a$ receives from all the other variables attached to
it. In order to clarify this point, we refer to Fig.~\ref{fig:factor}
(right), that is we focus on a small portion of the graph. As the graph is
locally tree-like, the variables $\s_i$, $i=1,2,...,K-1$, are only connected
through clause $a$ if $N$ is large enough. If $a$ is absent, the total
energy of the system can be written as $E^{N}(\s_1,...\s_{K-1}) =
A-\sum_{i=1}^{K-1} h_i\s_i$. We have used here the assumption that the
probability of these $\s_i$ factorizes. This is again motivated by the local
tree structure, but we shall go back to this point. After clause $a$ is
added, $E^{N+1}(\s_0,\s_1,...\s_{K-1}) = E^{N}(\s_1,...\s_{K-1}) +
\eps_a(\s_0,\s_1,...\s_{K-1})$ and the variables rearrange in order to
minimize the total cost. The minimization then defines $\tilde\eps(\s_0)$
from
\begin{equation}
  A-\sum_{i=1}^{K-1}|h_i|+\tilde\eps(\s_0) 
  = 
  \min_{\s_1,\ldots\s_{K-1}} E^{N+1}(\s_0,\s_1,...\s_{K-1}) \ .
  \label{eq:mini}
\end{equation}
This $\tilde\eps(\s_0)$ is then the cost to be paid for adding one variable
with a fixed value $\s_0$. Without losing generality, it can be written as
\begin{equation}
  \tilde\eps(\s_0) \equiv \Delta_{a\to 0} - \s_0 u_{a\to 0} \ ,
\end{equation}
where $u_{a\to 0}$ is the cavity bias acting on the new variable and
$\Delta_{a\to 0}$ is related to the actual energy shift by 
\begin{equation}
 \Delta E \equiv 
 \min_{\s_0,\ldots \s_{K-1}} \big[ E^{N+1}(\s_0,\s_1,\ldots \s_{K-1}) 
 - E^{N}(\s_1,\ldots \s_{K-1})\big] = \Delta_{a\to 0} -|u_{a\to 0}| \ .
 \label{eq:delta}
\end{equation}
Given that $\eps_a$ can be $0$ or $2$, depending on the set of fields
$h_1,\ldots h_{K-1}$ one has four possibilities:
\begin{eqnarray}
  \tilde\eps(+1)=0\quad\textrm{and}\quad\tilde\eps(-1)=0 & \Rightarrow & 
   u=0 \ , \ \;\,\Delta=0 \,\label{eq:1}\\
  \tilde\eps(+1)=0\quad\textrm{and}\quad\tilde\eps(-1)=2 & \Rightarrow & 
   u=+1 \ , \Delta=1 \,\label{eq:2}\\
  \tilde\eps(+1)=2\quad\textrm{and}\quad\tilde\eps(-1)=0 & \Rightarrow & 
   u=-1\ , \Delta=1 \, \label{eq:3} \\
  \tilde\eps(+1)=2\quad\textrm{and}\quad\tilde\eps(-1)=2 & \Rightarrow & 
   u=0 \ , \ \;\,\Delta=2 \ . \label{eq:4}
\end{eqnarray}
In other words, a non-zero message $u$ is sent from clause $a$ to variable
$\s_0$ only if the satisfiability of clause $a$ depends on $\s_0$. A null
message ($u=0$) can occur in the two distinct cases (\ref{eq:1}) and
(\ref{eq:4}).

The main hypothesis we have done so far consisted in assuming that the two
variables are uncorrelated if they are distant (the energy is linear in the
$\s_i$'s). It turns out that, in a large region of the parameter
space~\cite{mepaze,biroli}, including the regime we are interested in, this
is not correct. This is due to the fact that the space of solutions breaks
into many disconnected components if $\alpha$ is greater than a critical
value. In order to deal with this case, one has to introduce, for each
directed link, a probability distribution of the cavity biases, namely ${\sf
q}(u) \equiv \eta^+ \delta_{u,+1} + \eta^- \delta_{u,-1} + (1-\eta^+-\eta^-)
\delta_{u,0}$. The hypothesis of no correlation holds if the phase space is
restricted to one component. The interpretation of ${\sf q}_{a\to i}(u_{a\to
i})$ is the probability that a cavity bias $u_{a\to i}$ is sent from clause
$a$ to variable $i$ when one component is picked at
random~\cite{meze}. According to the rules
(\ref{eq:1},\ref{eq:2},\ref{eq:3},\ref{eq:4}), and with the topology of
Fig.~\ref{fig:factor} (right) as a reference for notations, the Survey
Propagation (SP, \cite{sp}) equations are then
\begin{eqnarray}
  \eta^+_{a\to 0} & \propto & \textrm{Prob} \left[
  \left\{ \left(u_{b\to i_1}\right)_{b\in i_1\backslash a },
  \ldots
  \left(u_{b\to i_{K-1}}\right)_{b\in i_{K-1}\backslash a } \right\}
  \bigg| ~\left(\tilde\eps_a(+1)<\tilde\eps_a(-1)\right) 
  \right] e^{-y\Delta E}\ ,
  \label{eq:etap}\\
  \eta^-_{a\to 0} & \propto & \textrm{Prob}  \left[
  \left\{ \left(u_{b\to i_1}\right)_{b\in i_1\backslash a },
  \ldots
  \left(u_{b\to i_{K-1}}\right)_{b\in i_{K-1}\backslash a } \right\}
  \bigg|~\left( \tilde\eps_a(+1)> \tilde\eps_a(-1)\right)
  \right] e^{-y\Delta E}\ ,
  \label{eq:etam}\\
  \eta^0_{a\to 0} & \propto & \textrm{Prob} \left [
  \left\{ \left(u_{b\to i_1}\right)_{b\in i_1\backslash a },
  \ldots
  \left(u_{b\to i_{K-1}}\right)_{b\in i_{K-1}\backslash a } \right\}
  \bigg|~\left( \tilde\eps_a(+1)=\tilde\eps_a(+1)\right)
  \right] e^{-y\Delta E} \ ,
   \label{eq:eta0}  
\end{eqnarray}
where the energy shift $\Delta E$ is given in eq.~(\ref{eq:delta}) and is
non-zero only when the constraint is UNSAT for any value of $\s_0$
(cf. eq.~(\ref{eq:4})). The crucial reweighting term exp$(-y\Delta E)$ thus
acts as a ``penalty'' factor each time a clause can not be satisfied. This
term is necessary in the UNSAT regime which we explore here (while simpler
equations with $y=\infty$ are enough to study the SAT phase). 

\subsection{ Encoding: SP and decimation}

For each fixed value of $y$, the iterative solution for the SP
equations can be implemented on a single sample~\cite{sp}, {\it i.e.}
on a given graph where we know all the function nodes involved. The
cavity probability distributions ${\sf q}(u)$'s are updated by picking
up one edge at random and using
(\ref{eq:1},\ref{eq:2},\ref{eq:3},\ref{eq:4}). This procedure is
iterated until convergence. This yields a set of messages
$\{\eta_{a\to i}^-, \eta_{a\to i}^0, \eta_{a\to i}^+\}$ on each edge
of the factor graph which is the solution of the SP equations. This
solution provides very useful information about the single instance
that can be used for decimation. As explained in~\cite{spy}, the
finite value of $y$ used for this purpose must be properly chosen.
Given the solution for this $y$, one can compute the distribution
$P(H_i)$ of the total bias $H_i=\sum_{a\in i} u_{a\to i}$ on each
variable. One can then fix the most biased variables, {\it i.e.} the
one with the largest $|P(H_i=+1)-P(H_i=-1)|$, to the value suggested
by the $P(H_i)$ itself. This leads to a reduced problem with $N-1$
variables. After solving again the SP equations for the reduced
problem, the new most-biased variable is fixed and one goes on until
the problem is reduced to an ``easy'' instance. This can be finally
solved by some conventional heuristic ({\it e.g.} walksat or simulated
annealing). A significant improvement of the decimation performance
can be obtained by using a backtracking
procedure~\cite{spy,giorgioback}: At each step we also rank the fixed
variables with a strongly opposed bias and unfix the most ``unstable''
variable with finite probability. The algorithm described here is
called Survey Inspired Decimation (SID) and its peculiar versions have
been shown to be very useful in many CSP problems recently.
Unfortunately, the basic version of SID does not work for the XORSAT
problem because of the symmetric character of the function nodes that
is reflected in a large number of unbiased variables (some
improvements appear to be possible \cite{private}).

\subsection{Statistical analysis: Theoretical performance}

One can perform a statistical analysis of the solutions of the SP equations
by population dynamics~\cite{meze}. The knowledge of the function node
$\eps_a$ allows to build up a table of values of $u$ for each configuration
of the local fields $h_1,\ldots,h_{K-1}$ (this is done according to the
minimization procedure described above). We then start with some initial
(random) population of $\vec\eta_i\equiv\{\eta^+_i,\eta^0_i,\eta^-_i\}$, for
$i=1,\ldots N$.  We extract a Poisson number $p$ of neighbors and $p$
probabilities $\vec\eta_{i_1},\dots \vec\eta_{i_p}$. According to these
weights, $p$ biases $u_1,\ldots u_p$ are generated and their sum computed,
$h_i=\sum_{a=1}^p u_{a}$. Once we have $K-1$ of these fields we perform the
minimization in (\ref{eq:mini}) and compute the new probability for $u_{a}$
according to the rules~(\ref{eq:1},\ref{eq:2},\ref{eq:3},\ref{eq:4}). The
whole process is then iterated until a stationary distribution of the cavity
biases is reached.  This method for solving the SP equations is very
flexible with respect to the change of the choice of the node, because this
choice just enters in the calculation once at the beginning of the algorithm
in order to initialize some tables.  One can also study problems with many
different types of nodes in a given problem: in this case, one among them is
randomly chosen each time the updating is performed. We finally stress
that once the probability distribution of the cavity biases is known the
ground state energy of the problem can be computed according to the
formalism introduced in~\cite{meze}. The expression for the energy in term
of the probability distributions as
\begin{eqnarray}
  \label{eq:energy}
  E(\alpha) &=& \max_{y} \Phi(\alpha,y)\ ,\\
  \Phi(\alpha,y) &=& 
  \frac{(-1)}{y}\left\{ [1+(K-1)\alpha] \overline{\log A_p(y)}
  -(K-1)\alpha \overline{\log A_{p+1}(y)} \right\} \ ,\\
  A_p(y) & \equiv & \sum_{u_1,\ldots u_p} {\sf q}_1(u_1) \cdots {\sf q}_p(u_p) 
 \exp\left( y\big|\sum_{a=1}^p u_a\big | -y \sum_{a=1}^p| u_a|\right) \ ,
\end{eqnarray}
the overline standing for an average over the Poisson distribution and the
choice of the distributions ${\sf q}_1,\ldots{\sf q}_p$ in the population.
Recalling that $R=1/\alpha$, the average distortion ({\sl i.e.}, the
theoretical capacity) of a compressor based on this CSP is computed through
equations (\ref{eq:distortion}) and (\ref{eq:energy}).

\begin{figure}[t]
  \psfrag{(a,y)}[][][1.9]{$(\alpha,y)$}
  \psfrag{y}[][][1.9]{$y$}
  \psfrag{2220}[][][1.7]{$\{2,2,2,0\}$}
  \psfrag{0220}[][][1.7]{$\{0,2,2,0\}$}
  \psfrag{2020}[][][1.7]{$\{2,0,2,0\}$}
  \psfrag{0020}[][][1.7]{$\{0,0,2,0\}$}
  \psfrag{2200}[][][1.7]{$\{2,2,0,0\}$}
  \psfrag{0200}[][][1.7]{$\{0,2,0,0\}$}
  \centering
  \includegraphics[angle=-90,width=.75\columnwidth]{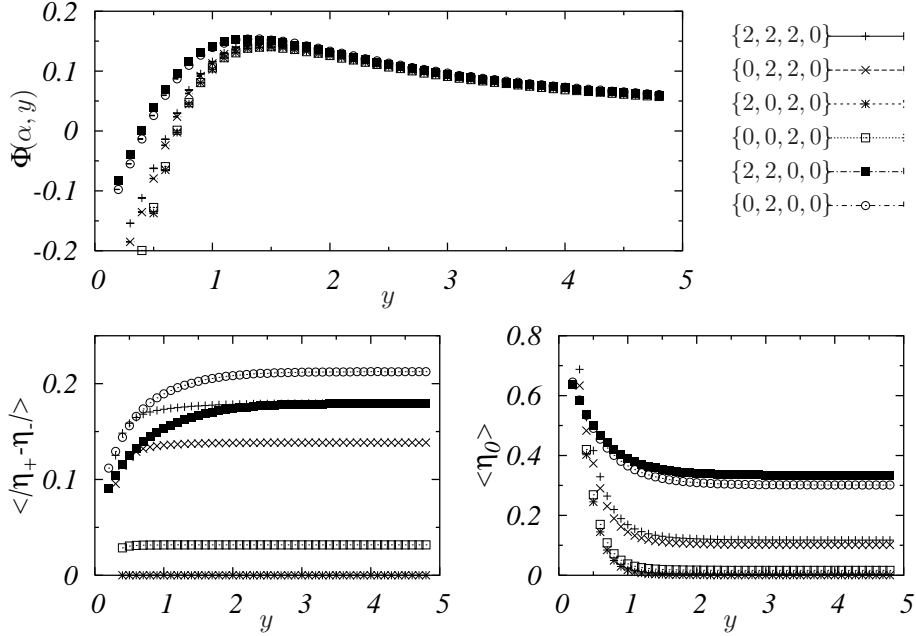}
  \caption{Top: The free energy $\Phi(\alpha,y)$ of a number of symmetric
    nodes with $K=7$ at $\alpha=1.4$, classified according to the rule given
    in the text. Bottom:  The asymmetry $|\eta_+-\eta_-|$ of these nodes
    (left) is measured as well as the ``paramagnetic degree"
    $\eta_0=1-\eta_--\eta_+$ (right). Note that the usual 7-XOR node
    corresponds to the $\{2,0,2,0\}$ case in this notation.}
  \label{fig:k7}
\end{figure}

Let us study here as an example a family of function nodes whose energy is
fully invariant under permutations of the arguments. These nodes can be
classified according to the energy of the node for a given value of the
``magnetization'' $m\equiv\sum_{i=1}^K \s_i$.  We keep to odd $K$ and label
a particular node by the sequence of values
$\{\eps(m=K),\eps(m=K-2),\ldots,\eps(m=1)\}$. In Fig.~\ref{fig:k7} we report
the results of the population dynamics algorithm for some types of nodes at
$K=7$. According to eq.~(\ref{eq:energy}), the ground state energy
$E(\alpha)$ corresponds to the maximum over $y$ of the free energy
$\Phi(\alpha,y)$ represented in the top plot. The theoretical performance of
these nodes are quite close to the PSC case (the XOR node, characterized by
the sequence $\{2,0,2,0\}$).

\section{Non-linear nodes}

We consider in this section the function nodes that give the best
performance for compression, both form the theoretical point of view
(theoretical capacity close the parity-check nodes) and from the algorithmic
aspect (the SID algorithm at finite $y$ is found to converge in the UNSAT
regime, thus giving an explicit encoding algorithm).  These non-linear
function nodes are defined as follows. We recall that the output
$\eps_{XOR}$ of a $K$-XORSAT node is given by twice the sum modulo 2 of the
input bits. We label each configuration $\vec \s \in \{0,1\}^K$ with an
integer $l\in[1,2^K]$ and consider a random permutation $\pi$ of the vector
$\{1,2,\ldots,2^K\}$. Then, we can associate the output of the {\em random
node} by letting $\eps^{(\pi)}(l) = \eps_{XOR}(\pi(l))$. In this way we are
left with a random but balanced output which can be different from
XOR. Also, it is clear that they are not more defined by a linear formula
over the boolean variables.

\begin{figure}
  \centering
  \includegraphics[angle=-90,width=.8\columnwidth]{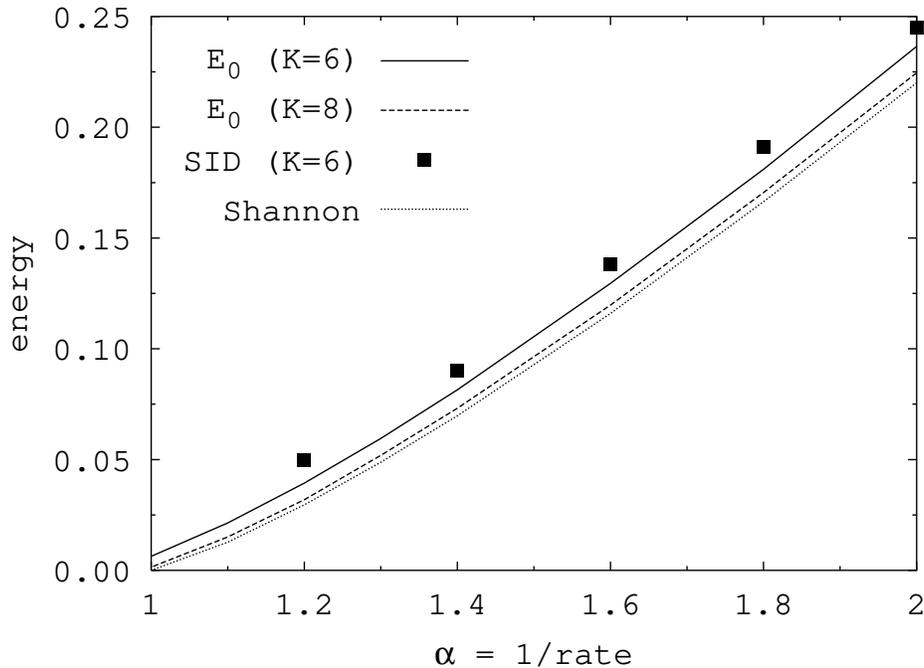}
  \caption{The ground state energy for a system with $30$ non-linear
  nodes. We also plot the Shannon bound and the performance of the SID
  algorithm for the $K=6$ case.}
  \label{fig:rncfr}
\end{figure}

We can take advantage of the formalism introduced in the previous section in
order to study the theoretical performance of these new function nodes. In
particular, we have used $10$ to $30$ different random nodes to build the
factor graphs. In order to improve the performance, we have also forbidden
``fully-canalizing'' nodes, that is nodes whose SAT character depends on
just one variable.

In Fig.~\ref{fig:rncfr} we show our results. The ground state energy is
shown to quickly approach the Shannon bound as $K$ increases and for any
$\alpha$. As an example, at $\alpha=2$ (corresponding to a compression rate
$R=1/2$) the difference between the $K=8$ value and the theoretical limit is
$\simeq 2\%$. This looks very promising from the point of view of data
compression. Furthermore, the results obtained from the SID (same plot) show
that in this case the algorithm does converge and its performance is very
good. It should be noted that when $K$ becomes large the difference between
SP and the Belief Propagation (BP) becomes small (this can be seen for
instance in the analysis of~\cite{cime}), so in fact BP does also provide a
good encoding algorithm for $K\gtrsim 6$. At fixed $K$, the time needed to
solve the SP equations is $\mathcal{O}(N\log N)$. The actual computational
time required by the decimation process is of the same order and it slightly
depends on the details of the SID algorithm ({\it e.g.}  the number of
variables fixed at each iteration, whether or not a backtracking procedure
is used, which kind of heuristic is adopted to solve an ``easy'' reduced
instance, the proper definition of the latter, etc...). The dependence on
$K$ is exponential.  Thus, even if increasing $K$ is good from the
theoretical point of view, it turns out to be very difficult to work at high
$K$. In practice, it takes a few hours to compress a string of $N=1000$
bits at $K=6$ by using our general purpose software. We think that some more
specified code would lead to a better performance.

To conclude, we have shown how the methods of statistical mechanics,
properly adapted to deal with a new class of constraint satisfaction
problems, allow to implement a new protocol for data compression. The new
tool introduced here, the non-linear gates, looks very promising for other
practical applications in information theory.

\subsection*{Acknowledgments}

We thank A.~Montanari, F.~Ricci-Tersenghi, D.~Saad, M.~Wainwright and
J.~S.~Yedidia for interesting discussions. S.~C. is supported by EC through
the network MTR 2002-00307, DYGLAGEMEM. This work has been supported in part
by the EC through the network MTR 2002-00319 STIPCO and the FP6 IST
consortium EVERGROW.


\end{document}